# Direct observation of the perpendicular shape anisotropy and thermal stability of p-STT-MRAM nano-pillars

Trevor P. Almeida[1,2]*, Steven Lequeux[3], Alvaro Palomino[3], Ricardo C. Sousa[3], Olivier Fruchart[3], Ioan Lucian Prejbeanu[3], Bernard Dieny[3], Aurélien Masseboeuf[3] and David Cooper[1]

[1]Univ. Grenoble Alpes, CEA, Leti, F-38000 Grenoble, France.
[2]SUPA, School of Physics and Astronomy, University of Glasgow, G12 8QQ, UK.
[3]Univ. Grenoble Alpes, CEA, CNRS, Grenoble INP, SPINTEC, 38000 Grenoble, France.

Perpendicular shape anisotropy (PSA) offers a practical solution to downscale spin-transfer torque Magnetic Random-Access Memory (STT-MRAM) beyond the sub-20 nm technology node whilst retaining thermal stability of the storage layer magnetization. However, our understanding of the thermomagnetic behavior of PSA-STT-MRAM is often indirect, relying on magnetoresistance measurements and micromagnetic modelling. Here, the magnetism of a FeCoB / NiFe PSA-STT-MRAM nano-pillar is investigated using off-axis electron holography, providing spatially resolved magnetic information as a function of temperature, which has been previously inaccessible. Magnetic induction maps reveal the micromagnetic configuration of the NiFe storage layer (~ 60 nm high, ≤ 20 nm diameter), confirming the PSA induced by its 3:1 aspect ratio. *In-situ* heating demonstrates that the PSA of the FeCoB / NiFe composite storage layer is maintained up to at least 250 $^{0}$C, and direct quantitative measurements reveal the very moderate decrease of magnetic induction with temperature. Hence, this study shows explicitly that PSA provides significant stability in STT-MRAM applications that require reliable performance over a range of operating temperatures.

*Corresponding author:
Tel: +44 (0) 141 330 4712
Email: trevor.almeida@glasgow.ac.uk

## 1. Introduction

Magnetic random-access memory (MRAM) is a non-volatile memory based on the storage of one bit of information by a ferromagnetic memory cell. The discovery of the spin-transfer torque (STT) effect has made MRAM industrially relevant, thanks to the ability to write cells with an electric current. STT-MRAM can be easily integrated with CMOS technology[6], has low energy consumption[7], superior endurance[7,8], and rather high areal density[9]. In practice, STT-MRAM involves the use of a magnetic tunnel junction (MTJ) comprising an MgO tunnel barrier (1 – 1.5 nm thick) separating two thin perpendicularly-magnetized layers, one magnetically-pinned reference layer, and one switchable storage layer. This design is called p-STT-MRAM, the perpendicular anisotropy resulting from interfacial electronic hybridization effects occurring at the interface between the magnetic electrodes and the oxide tunnel barrier[10]. The electrical resistance of the MTJ changes significantly when the layers are magnetized in parallel and antiparallel states, providing a system of readable / writable '0' or '1' binary information. Increasing the areal bit density of p-STT-MRAM requires to reduce the in-plane size or diameter of the nano-patterned MTJ[11]. However, below a characteristic magnetic length scale around 20nm, this comes with an excessive decrease of the thermal stability of the magnetic moment of the ultrathin storage layer. One solution is to increase the storage layer thickness to larger than its diameter, so that its out-of-plane aspect ratio combined with large volume, provides additional thermal stability through perpendicular shape anisotropy (PSA) [12,13]. Previous studies have shown that the PSA-STT-MRAM are indeed highly thermally stable, making them a practical solution to downsize scalability of STT-MRAM at sub-20 nm technology nodes[14,15]. However, our knowledge of the thermal stability of these STT-MRAM nano-pillars is often indirect, relying on magnetoresistance measurements and micromagnetic modelling. In order to understand fully their thermomagnetic behavior, it is necessary to examine the effect of temperature directly. The advanced transmission electron microscopy (TEM) technique of off-axis electron holography allows imaging of magnetization within nano-scale materials, with high spatial resolution and sensitivity to induction field components transverse to the electron beam[16]. Combining electron holography with *in-situ* heating within the TEM has already allowed direct imaging of the thermal stability of nano-scale signal carriers and fields of magnetic minerals[17-19], meteorites[20] and pre-patterned MTJ conducting pillars[21].

In this paper, we use electron holography to image the micromagnetic configuration of an individual ≤ 20 nm diameter FeCoB / NiFe STT-MRAM nano-pillar and acquire quantitative measurements of magnetic induction from the high-aspect-ratio FeCoB / NiFe composite free layer. In addition, we experimentally demonstrate the influence of PSA on its thermal stability through *in-situ* heating within the TEM to 250 $^0$C. This study provides direct evidence of the PSA exhibited by the high-aspect-ratio free layer, and its improved thermal stability compared to standard p-STT-MRAM.

## 2. Experimental

An array of FeCoB / NiFe PSA-STT-MRAM nano-pillars was fabricated through sequential e-beam lithography, reactive ion etching and ion beam etching[12]. The full stack of PSA MTJs has the following composition: $SiO_2$ / Pt (25 nm) / SAF / Ta (0.3 nm) / FeCoB (1.1 nm) / MgO (1.2 nm) / FeCoB (1.4 nm) / W (0.2 nm) / NiFe (60 nm) / Ta (1 nm) / Ru (3 nm) / Ta (150 nm), where the thick NiFe layer provides the PSA[12,14]. SAF stands for a synthetic antiferromagnet, providing a compensated-moment antiferromagnetically-pinned magnetic reference as the bottom layer. A protective layer of organic resin (~ 1 µm) was spin-coated onto the array prior to being inserted into a Thermo Fisher Strata 400S dual-beam focused ion beam (FIB) / secondary electron microscope (SEM) for TEM sample preparation. A protective W layer was deposited over the resin and trenches were FIB irradiated until a ~ 3 µm × ~ 20 µm lamella was prepared and transferred to Omniprobe copper lift-out grids, where it was thinned to a thickness of ~ 300 nm using conventional FIB methods. The protective resist layer was etched away using a plasma cleaner and the remaining W layer was broken off with the micromanipulator. High angle annular dark field (HAADF) scanning TEM (STEM) imaging was performed using a probe-$C_S$-corrected Thermo Fisher Titan TEM at 200kV. Corresponding chemical analysis was provided by energy dispersive X-ray (EDX) spectroscopy using Bruker SDD Super X Fast EDX four-detectors. Off-axis electron holograms were acquired under field-free conditions in Lorentz mode on a Gatan OneView 4K camera using a Thermo Fisher Titan TEM equipped with an image-$C_S$ corrector and an electron biprism. A voltage of 220 V was applied to the biprism, resulting in an interference fringe spacing of ~ 1.7 nm. Stacks of 8 electron holograms (each acquired for 4s) were aligned and then averaged using the Holoview software[22] to improve the signal to noise ratio of the reconstructed phase images. The magnetization states of the nano-pillars were visualized through separation of the magnetic contribution to the phase shift from the mean inner potential (MIP), achieved by tilting to ± 40$^0$ and applying the strong field of the objective lens (< 1.5 T) to reverse the magnetic contribution[23]. *In-situ* heating up to 250°C was performed using a Gatan heating holder under field-free magnetic conditions, allowing 30 minutes to stabilize from thermal drift at each temperature interval. The heating was repeated and the magnetic reversal was performed at each temperature interval to isolate the MIP, and subtracted from the first heating to reconstruct the thermomagnetic behavior of the nano-pillars[18,19]. For the construction of magnetic induction maps, the cosine of the magnetic contribution to the phase shift was amplified ($\times$ 150) to produce magnetic phase contours and arrows were added to show the direction of the projected induction.

## 3. Results

Figure 1 presents the chemical composition, morphology and structure of an individual nano-pillar from the etched FeCoB / NiFe stack. The HAADF STEM image (Fig. 1a) displays the general morphology of the nano-pillar with a distinct difference in Z-contrast between the top Ta mask and lower shaft. This is confirmed by EDX chemical mapping of Fig. 1b, revealing the NiFe section of the nano-pillar is ~ 60 nm high with a diameter of ≤ 20 nm, and is separated from the hard Ta mask by the ~ 3 nm Ru layer. The O signal is dispersed across the entire nano-pillar and its physical state is explored through precession diffraction acquired from the boxed region (red) in the EDX chemical map of Fig. 1c. The weakly crystalline / amorphous state of the surface of the nano-pillar highlighted in the scanned area of Fig. 1d (left) is demonstrated in the associated electron diffraction image (right). Conversely, Fig. 1e shows the crystalline state of the center of the nano-pillar in the scanned area (left) and associated electron diffraction (right). The line profile of Fig. 1f is acquired from the red arrow in Fig. 1c (averaged vertically over ~15 nm) and displays the cross-section of normalized x-rays % for Fe, Ni and O content. The Ni and Fe profiles overlap consistently and the diffraction data of Fig. 1e demonstrates a homogeneous, crystalline NiFe core of the nano-pillar with measured composition of 81 at% Ni and 19 at% Fe. The O-rich phase is confined to the surface and considering its weakly crystalline / amorphous state, is largely attributed to residual organic resin from the preparation process.

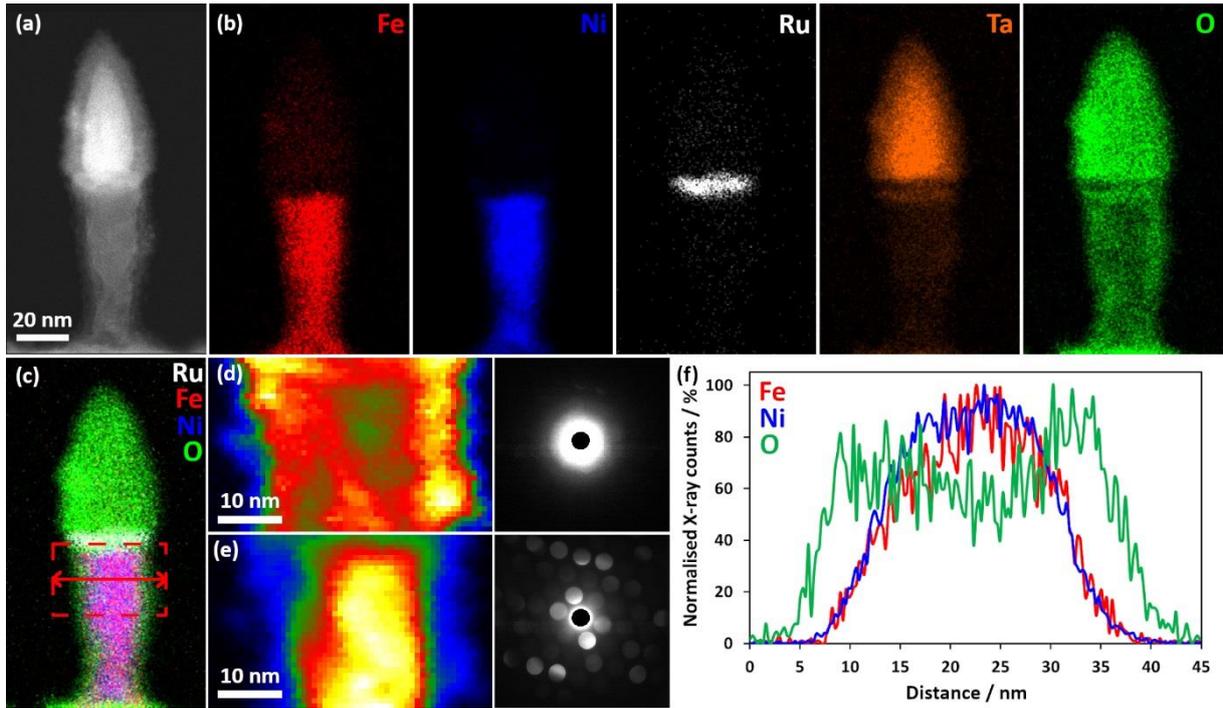

**Figure 1. Overview of the morphology, chemical composition and structural properties of the nano-pillar.** (a) HAADF STEM image of the nano-pillar; and (b) associated EDX chemical maps showing the elemental distribution of iron, nickel, ruthenium, tantalum and oxygen. (c) EDX chemical map highlighting the iron / nickel-rich center and oxygen-rich surface of the nano-pillar. (d,e) Precession electron diffraction acquired from the boxed region (red) in (c), showing the (d) amorphous sidewall surface; and (e) crystalline core. (f) Elemental line profile acquired along the red arrow in (c) and averaged vertically over ~ 15 nm, displaying the cross-section of normalized x-rays % for iron, nickel and oxygen content.

Figure 2 presents phase information reconstructed from electron holography of the nano-pillar, including quantification of magnetic induction and associated magnetization. As a guide, the MIP contribution (Fig. 2a) shows the morphology of the nano-pillar at the same magnification as the associated projected magnetic induction map (Fig. 2b), which clearly demonstrates the existence of a magnetic dipole within the nano-pillar. This is emphasized by superimposing the associated magnetic induction over the EDX chemical map (Fig. 2c), where the magnetic contours flow along the major axis of the NiFe section. A line profile of the magnetic phase contribution ($\phi_m$) acquired along the white arrow in Fig. 2c (averaged vertically over ~15 nm) is overlaid over the chemical profile from the same region (Fig. 2d). The magnetic phase shift $\Delta\phi_m$ in the pillar can be used to quantify the integrated in-plane magnetic induction related to the nano-pillar, where the minimum and maximum $\phi_m$ values [Fig. 2c, dotted vertical lines (pink)] coincide with the full width half maximum of the Ni / Fe chemical profiles. Extracting quantitative calculations of magnetization from such local observations must be handled with care as it displays integrated information (see Supplementary materials for details on the calculations summarized below). At first approximation, the nano-pillar can be treated as a uniformly-magnetized infinite cylinder of radius *r*. The magnetic induction transverse to the electron beam direction is considered to arise from magnetization only (neglecting dipolars fields, i.e., stray field and internal demagnetizing field *i.e.*) and can be estimated with the Aharonov-Bohm effect using a radius of ~ 8.6 nm (obtained from Fig. 2d). A first value of $\mu_0 M_s$ = 0.82 ± 0.06 T is extracted with the standard deviation determined from the noise of free space. To better quantify this important parameter, we can estimate the associated demagnetizing field of such a structure with the $F_{ijk}$ functions from Hubert's formalism using a prism approach[24], at the exact same location of the measurement of the integrated magnetic flux obtained with the phase slope. We then obtain a better estimation of $\mu_0 M_s$ = 1.19 ± 0.08 T with the removal of the integrated demagnetizing and stray field globally antiparallel to magnetization, which lowered our first approximation. Subsequently, to overcome the geometrical approximation for the integrated demagnetizing field estimation with a prismatic geometry, we performed a tomographic reconstruction through inverse Abel transformation[25] to extract the magnetic flux density in the core vicinity of the cylinder. The resulting value was finally adjusted with the local estimation of demagnetizing field through $F_{ijk}$ functions, as the tomography method helps to deconvolve the B field but not the internal $H_d$. Our final extraction leads to a value for the spontaneous magnetization induction of the NiFe nano-pillar of **$\mu_0 M_s$ = 1.14 ± 0.01 T**. This value is in reasonable agreement with the spontaneous induction for a $Ni_{81}Fe_{19}$ chemical composition, expected to reach 1.05 T [26]. We ascribe the difference to uncertainties related to the exact value of the composition of the alloy the diameter of the PSA, and the sharpness of its outer surface. Being systematic, these errors are expected to have no impact on the thermomagnetic measurements reported in the following. More details in our analysis of this quantification are given in the supplementary materials.

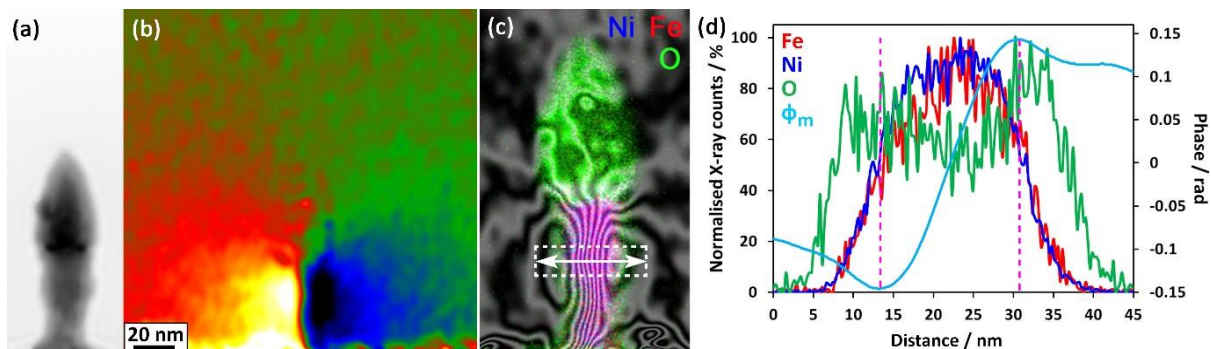

**Figure 2. Overview of the magnetic induction associated with the nano-pillar and its quantification.** (a,b) Reconstructed (a) MIP; and (b) magnetic contribution to the total phase, revealing the PSA along the major axis of the nano-pillar. (c) Combined EDX / magnetic induction map showing magnetic contours flowing along the major axis of the NiFe section. The contour spacing is 0.042 rad (cosine amplified by 150 times). (d) Line profile acquired along the white arrow in (c) and averaged vertically over ~ 15 nm around the mid-height of the PSA free layer, superimposing the cross-section of magnetic phase over the chemical profile.

Figure 3 presents the thermomagnetic behavior of the nano-pillar during *in-situ* heating from room temperature to 250 $^0$C. The initial combined EDX / magnetic induction map of Fig. 3a displays the nano-pillar at 20 $^0$C, with the magnetization pointing from the bottom to the top of the NiFe free layer and confirming its PSA. The associated magnetic induction maps of Figs 3b-h display the same nano-pillar during heating from 100 $^0$C to 250 $^0$C at 25 $^0$C intervals. It is evident that the NiFe section retains its upward direction of magnetization at all temperatures up to 250 $^0$C. The thermomagnetic stability of the NiFe section in the nano-pillar is supported by Figure 4, which presents corresponding quantitative measurements of magnetic induction at each temperature interval. The $\Delta\phi_m$ was measured from the same area presented in Fig. 2c,d and $\vec{B}$ calculated using equation S2 (Supplementary information), which is proportional to magnetization. The average value of $\vec{B}$ is ~ 0.8 ± 0.06T and the trend-line exhibits a decrease of ~0.08T between 20 $^0$C and 250 $^0$C, which is in good agreement with the thermal behavior of bulk permalloy ($Ni_{80}Fe_{20}$)[27,28]. As the contribution of dipolar field to the measured induction field is proportional to magnetization for a PSA, we expected that Fig.4 is representative of the thermal decay of magnetization in the PSA.

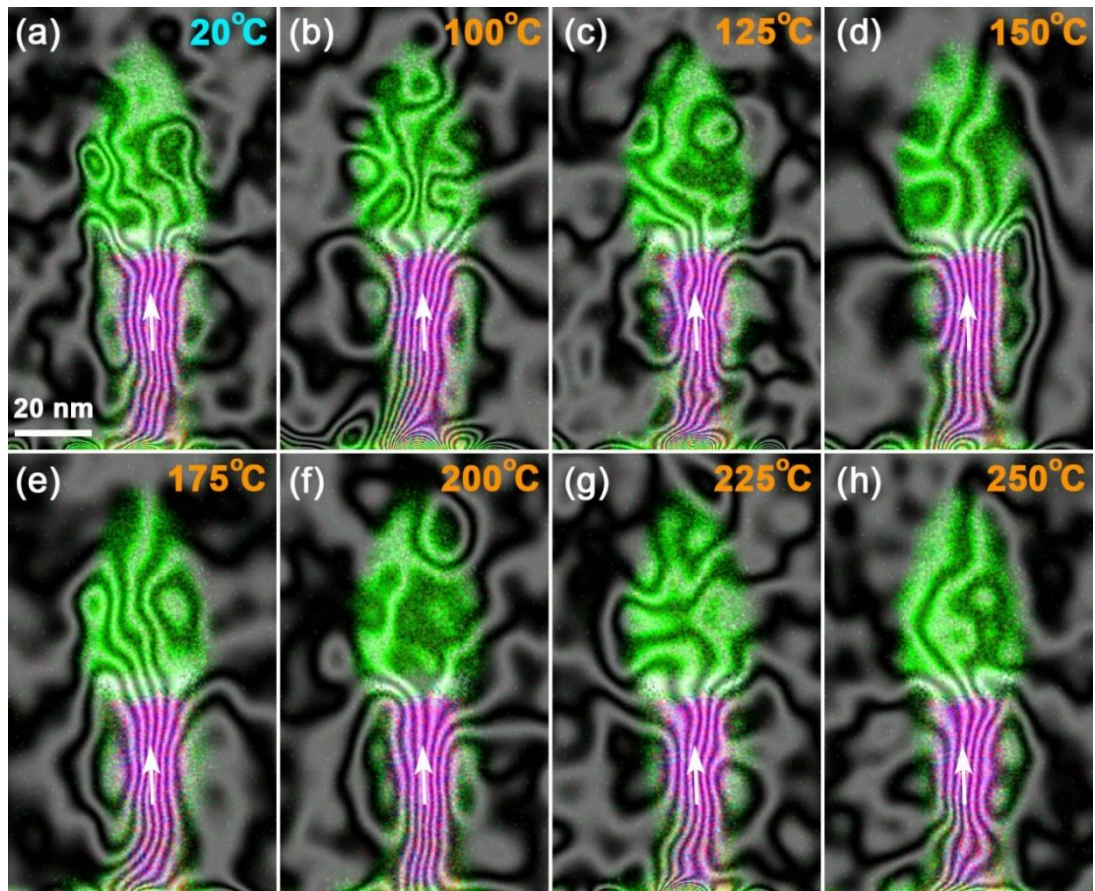

**Figure 3. Visualization of the thermomagnetic stability within the nano-pillar.** (a-h) Combined EDX / magnetic induction maps showing the PSA in the NiFe section of the nano-pillar acquired at (a) 20 °C; or during *in-situ* heating at 25 °C intervals from (b) 100 °C to (h) 250 °C. The contour spacing is 0.042 rad for all the magnetic induction maps (cosine multiplied by 150 times) and the magnetization direction is shown using arrows.

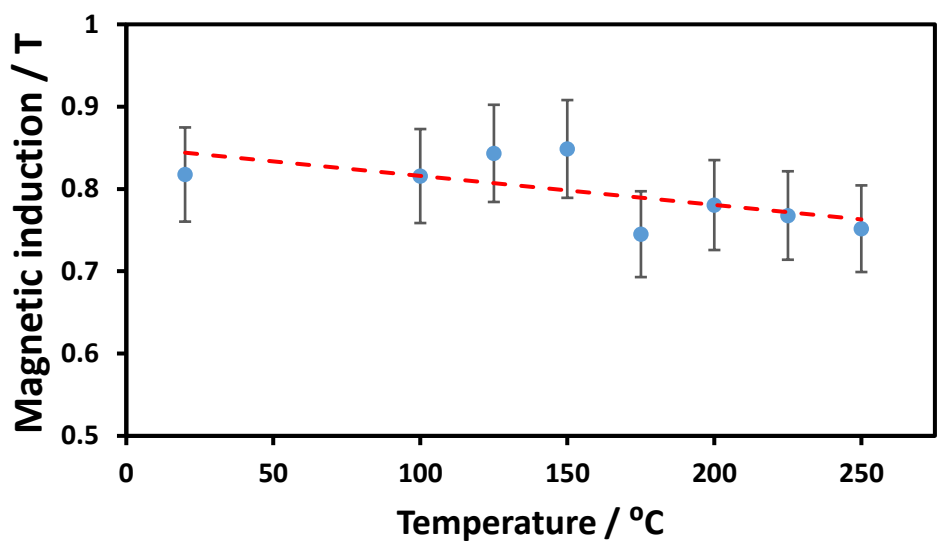

**Figure 4. Magnetic induction associated to the nano-pillar as a function of temperature.** Calculated magnetic induction transverse to the electron beam using the direct measurement of magnetic contribution to the phase from the NiFe section shown in Figure 3 with temperature, i.e., including a contribution from the demagnetizing and stray fields globally opposite to magnetization in the pillar, from 20 °C to 250 °C.

## 4. Discussion

The main reason to fabricate STT-MRAM nano-pillars with PSA is to stabilize and preserve the magnetic configuration of the free FeCoB / NiFe composite layer in response to temperature fluctuations during operation. It is evident from the *in-situ* heating TEM experiment that the dipolar magnetic orientation along the major axis of the NiFe section (bottom to top) is retained at all temperature intervals up to at least 250 $^0$C. In addition, the associated direct measurements of $\vec{B}$ (average ≈ 0.8T ± 0.06T) did not vary sufficiently to suggest a change in magnetic orientation, nor magnetization. The slight decreasing trend in magnitude of $\vec{B}$ is attributed to an expected reduction in spontaneous magnetization when heating towards the Curie temperature[29], as would be observed in the same bulk material. As the shape-anisotropy energy barrier of the nano-pillar scales linearly with $M_S^2$, this confirms that the energy barrier of switching between parallel and anti-parallel states has not been exceeded up to at least 250 $^0$C, and the PSA has provided significant thermal stability compared to standard p-STT-MRAM stacks based on ultrathin films[12-15]. In standard p-STT-MRAM, the spontaneous magnetization may indeed decrease by a few tens of percent at 250°C due to enhanced fluctuations in low-dimensions for the ultrathin films involved, and the energy barrier associated with perpendicular anisotropy may decrease even more as it relies on interfacial magnetic anisotropy, which suffers from enhanced thermal decay due to its two-dimensional nature[30-32]. Further, the maximum temperature of *in-situ* heating used (250 $^0$C) far exceeds the operating temperatures for a range of STT-MRAM device applications, including commercial (70 °C), industrial (85 °C), automotive (125 °C) and military (150 °C)[33]. It is also comparable to the temperature of reflow soldering used to attach chips to a printed circuit board (260 °C)[34]. This thermal stability is crucial for many FLASH-replacement applications in which the program code is loaded in the chip prior to reflow soldering.

In conclusion, this electron holography study has provided explicit evidence of the PSA in ≤ 20 nm diameter p-STT-MRAM nano-pillar; previously only measured indirectly by magnetoresistance measurements and micromagnetic modelling. The enhanced thermal stability of PSA-STT-MRAM, compared with standard p-STT MRAM based on these films, has been confirmed by both imaging and quantitative measurements from their magnetic configuration during *in-situ* heating. This supports the use of PSA-STT-MRAM in a variety of applications that require reliable performance over a range of operating temperatures.

**Funding**

CEA-Leti is a Carnot Institute. This work was supported by the French ANR via Carnot funding and the European Research Council (ERC MAGICAL, 664209).

# Direct observation of the perpendicular shape anisotropy and thermal stability of p-STT-MRAM nano-pillars

Trevor P. Almeida[1,2]*, Steven Lequeux[3], Alvaro Palomino[3], Ricardo C. Sousa[3], Olivier Fruchart[3], Ioan Lucian Prejbeanu[3], Bernard Dieny[3], Aurélien Masseboeuf[3] and David Cooper[1]

[1]Univ. Grenoble Alpes, CEA, Leti, F-38000 Grenoble, France.
[2]SUPA, School of Physics and Astronomy, University of Glasgow, G12 8QQ, UK.
[3]Univ. Grenoble Alpes, CEA, CNRS, Grenoble INP, SPINTEC, 38000 Grenoble, France.

## Supplementary information

*Corresponding author:
Tel: +44 (0) 141 330 4712
Email: trevor.almeida@glasgow.ac.uk

**This PDF file includes:**

S1: Dipolar field calculation.

S2: Quantitative phase calculation.

# S1: Dipolar field calculation

We used Hubert's formalism[1] for calculating the magnetostatic field $\mathbf{H}_d(\mathbf{r})$ arising from the pillar, also called demagnetizing field inside the pillar, and stray field outside the pillar. Estimating the magnetostatic field is crucial to extract magnetization $\mathbf{M}(\mathbf{r})$ from the experimental results, which pertain to the magnetic induction field $\mathbf{B} = \mu_0(\mathbf{H} + \mathbf{M})$.

Hubert's formalism relies on the so-called $F_{ijk}$ functions, related to the $i^{th}$, $j^{th}$ and $k^{th}$ integrals along $x$, $y$ and $z$, respectively, of the core function $V(\mathbf{r})=1/\mathbf{r}$, involved in the calculation of the magnetostatic potential.

In practice, this formalism requires the consideration of prisms, not cylinders. Hence, as an approximation we considered a prism with the same height and cross-section as the experimental pillar: square section with $a = b = 15.24$ nm such as $ab = \pi r^2$, and $c = 60$ nm (see Fig. S1a for a schematic of such prism). Magnetic quantities are calculated for a charged plate in the $(x,y)$ plane, associated with magnetization along the axis of the pillar :

- The magnetic potential is calculated with the $F_{110}$ function
- The vertical component of magnetic field $\mathbf{H}_{d,z}$ is calculated with the $F_{11-1}$ function
- The vertical component of magnetic field integrated along the beam $(y)$ is calculated with the $F_{12-1}$ function
- The vertical component of magnetic field, integrated along the beam and averaged over a strip of height $\Delta z$, is calculated with the $F_{120}$ function. Note that this quantity is directly related to the magnetic potential, averaged along the beam direction $y$.

As not all expressions for the above $F_{ijk}$ functions are mentioned in the book of Hubert and Schäfer[1], here is the full list:

$$F_{000}(x,y,z) = \frac{1}{\sqrt{x^2+y^2+z^2}},$$

with $\phi(x,y,z) = \frac{Q}{4\pi} F_{000}(x,y,z)$ the magnetostatic potential associated with the magnetic charge $Q$.

$$F_{110}(x,y,z) = yL_x + xL_y - P_z$$

$$F_{11-1}(x,y,z) = -\frac{P_z}{z}$$

$$F_{12-1}(x,y,z) = -zL_x - \frac{y}{z}P_z$$

$$F_{120}(x,y,z) = xyL_y + \frac{1}{2}(v-w)L_x - yP_z - \frac{1}{2}xr$$

with the following functions: $u = x^2$, $v = y^2$, $w = z^2$, $r = \sqrt{x^2 + y^2 + z^2}$, $L_x = \text{Atanh}(x/r)$, $P_x = x\,\text{Atan}\,(yz/xr)$, with $L_x = 0$ and $P_x = 0$ for $x = 0$ (and permutations for $y$ and $z$).

In practice, we then performed the calculation of the $y$-integrated $z$ component of the dipolar field, further averaged along a $z$ slice (with height 15 nm) around a given $z$ position, similar to the experimental average made on experimental extractions. This allows us to subtract the suitable value of magnetic field, part of the experimentally-measured induction from cross-sections such as in Fig. 2d of the main manuscript. This can be applied to the two procedures we implemented (*i.e.* using the phase slope or the tomographic method respectively - see next section) to estimate the real magnetization value. It is worth noticing that our estimation may be affected by two important parameters:

1) The magnitude of dipolar magnetic field is strongly dependant on the selected vertical position of the slice (see Fig. S1d). To extract the most suitable data from the electron holography images and to calculate reliable values of magnetization, we selected the maximum experimental values of integrated induction. However, such experimental measurements are influenced by angles of projection as they are only sensitive to the in-plane component of the magnetic induction, and the extracted values includes these negative contributions from the demagnetizing field. Hence we are limited to finding a local extremum of induction from a 2D projection of a 3D magnetic field and assume that our local maximum truly corresponds to the middle height of the pillar (and the minimum of dipolar field), keeping in mind that our measurement could be affected by a remaining angle in the experimental projection.

2) All values computed here are given in [M] for the dipolar field and [M].nm for the $y$-integrated dipolar field. The latter are converted to Tesla (T.nm) based on the assumption of spontaneous magnetization of 1.2 Tesla[2]. A more accurate estimation would require an iterative procedure to account for the new corrected approximation of the spontaneous magnetization to compute new values of the demagnetization field. However, for our estimation such iterations should only act at the mT level, below our experimental errors. That is, the computation given in the next section is given in T.nm, knowing that it remains an approximation.

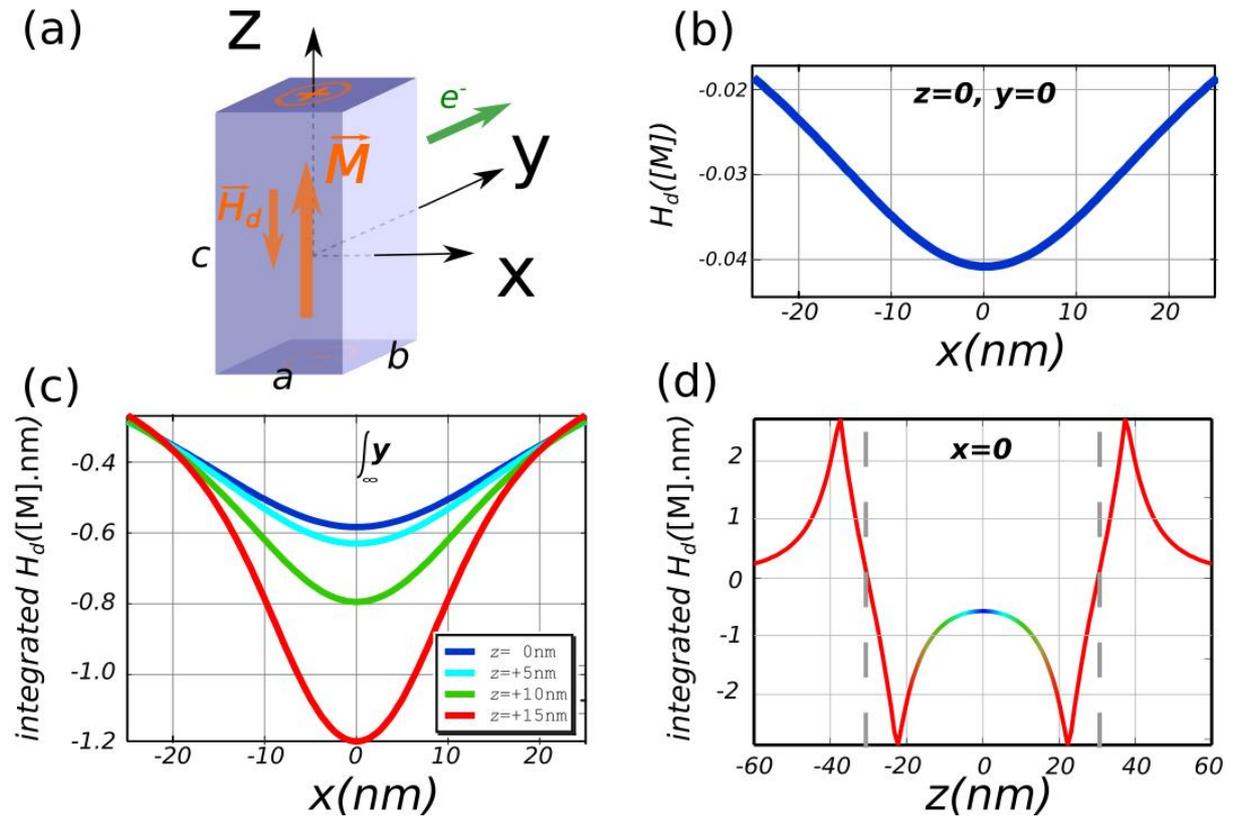

**Figure S1** (a) Scheme of the prism used for the calculation with corresponding values given in the text. The origin of coordinates is at the center of the prism (b) $x$ cross-section of the $z$ component of the demagnetizing field at mid-height of the prism and y=0, further averaged along a $z$ slice of height 15 nm to reproduce the experimental measurement. (c) $x$ cross-section of the $y$-integrated $z$ component of the demagnetizing field, further averaged along a $z$ slice of height 15 nm, calculated around several given $z$ positions from the center of the pillar ($z = 0$). (d) Profile along $z$ of the maximum in absolute value (always found at x=0, *i.e.* on the axis of the pillar) of the curves shown in (c), versus the position of the slice, the latter still with average along $z$ over 15 nm (the color code along $z$ is similar as in (c) for easy understanding).

## S2: Quantitative phase calculation

We used 3 different methods for quantifying the spontaneous magnetization in this study. These methods are details hereafter:

### Cylinder approximation:

In this first approximation, the nano-pillar is treated as a uniformly-magnetized infinite cylinder of radius $r$. In this case, the magnetic induction transverse to the electron beam direction is only arising from magnetization, and can be calculated using the following equation:

$$|\vec{B}| = \frac{\hbar \Delta \phi_m}{e \pi r^2} \qquad (S2)$$

where $e$ is the (positive) elementary charge, $\hbar$ is reduced Planck's constant and $\vec{B}$ is the in-plane magnetic induction[3]. Using the measured radius of $r \sim 8.6$ nm, the value of $\vec{B}$ is calculated as **0.82 ± 0.06 T** with the standard deviation determined from the noise of free space.

### Slope measurement and integrated demagnetizing field approximation:

A common method for determining a magnetic flux density from holography phase image is to use a phase profile fitting to extract the integrated flux in a linear part of its plot. The curve of Fig. 2b in the main manuscript has thus been fitted to obtain a slope of $30 \pm 2$ mrad.nm$^{-1}$. Such a slope has to be normalized with the Magnetic Flux Quantum

$$\Phi_0 = \frac{e}{2h} = 2.07 \ 10^{-15} \text{Wb} = 658 \ \text{T.nm}^2$$

obtain for the integrated magnetic flux corresponding to the measured phase shift : $19.7 \pm 1.4$ T.nm. This new estimation should give rise to a value of spontaneous magnetization of $1.14 \pm 0.08$ T, which does not include the influence of the demagnetizing field that would lower this value. However, such a measurement is a more precise extraction than the previous one which only relies on the overall phase shift, strongly influenced by lateral dipolar field, instead of the phase variation in the vicinity of the pillar.

Due to the finite length of the nano-pillar, it gives rise to a stray field and an internal demagnetizing field from the magnetic charges occurring at both pillar surfaces (see Fig. S1a). These fields contribute to the experimental measurement of the induction, overall opposite to the magnetization direction, so that the above value underestimates the spontaneous magnetization of the material. We thus used Hubert's formalism to compute the integrated demagnetizing field experienced by the electrons along their trajectory (see Fig. S1(a & d)): 0.72 T.nm is thus found at the middle of the pillar.

We can thus remove the underestimation of the measured integrated magnetic flux corresponding to the magnetization, leading to a final estimation of the spontaneous magnetization of **1.19 ± 0.08 T.** Nevertheless, the estimation of the integrated value of the demagnetizing field suffers from uncertainty as we selected its minimum value. This leads to a substantial underestimation if there is (i) any residual tilt in the experimental projection (S1, parameter 2; and Fig. S1d) or (ii) experimental geometry deviation to the model used that would lead to a displacement of such locale minimum. We eventually applied another correction using experimental estimation of external stray fields using a vectorial field "tomographic" reconstruction in the last section below.

### "Tomography" approach:

In order to take into account the influence of the stray field outside the nano-pillar, we reconstructed the 3D magnetic field from a single-phase image using symmetry arguments of the one-dimensional nature of the nano-pillar[4]. The reconstructed 3D field allows isolation of individual planes of magnetic induction through the center of the nanopillar (Fig. S2a) or plane behind the nano-pillar (Fig. S2b). The averaged line profile of $\vec{B}$ acquired from the arrow in Fig. S2a (white) is plotted in Fig. S2c, revealing the large positive value of $1.09 \pm 0.01$ T from the center of the nano-pillar. Two smaller negative values of 0.075 T are observed on either side of the central peak, and are similar in value to the averaged line profile acquired from the plane behind the nano-pillar (white arrow in Fig. S2b).

Such reconstruction enables the removal of the influence of external stray field on the electron wave as it travels past the nano-pillar. However, we still have to account for the internal demagnetizing field that was computed in previous section. For this, we use the estimation of the demagnetizing field in the vicinity of the pillar (Fig. S1b) which has been estimated at 49mT, leading to a corrected value of the saturated magnetization of **1.14 ± 0.01 T.** This value is in the same range of the one computed with the slope approximation corrected with the integral of the stray field. However, such value is substantially more precise and greater confidence is placed on the stray field calculation, especially considering the geometric approximation for the pillar is only estimated at the right center of the structure.

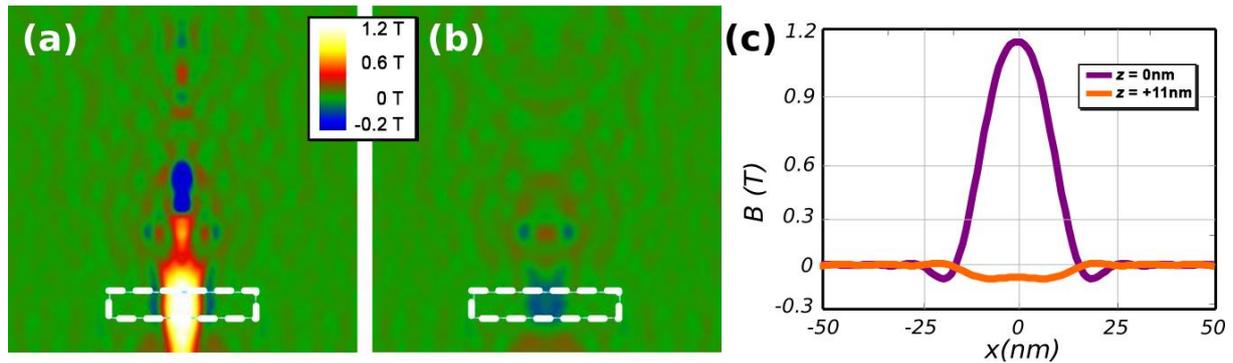

**Figure S2.** (a,b) Planes of magnetic induction acquired from a 3D reconstruction and measured at the (a) center (0nm); and (b) behind (11 nm) the nano-pillar. (f) Averaged line profiles of magnetic induction measured along the pillar (white dashed area in a & b) comparing the magnetic induction through the center and behind the nano-pillar.